\documentclass[prl,onecolumn,showpacs,preprint,amsmath,amssymb]{revtex4}

\usepackage{graphicx}

\begin{document}
\title{Weak localization in graphene flakes}
\author{F.~V.~Tikhonenko}
\author{D.~W.~Horsell}
\author{R.~V.~Gorbachev}
\author{A.~K.~Savchenko}
\affiliation{School of Physics, University of Exeter, Stocker Road, Exeter, EX4 4QL, U.K.}

\pacs{73.23.-b, 72.15.Rn, 73.43.Qt}

\begin{abstract}
We show that the manifestation of quantum interference in graphene
is very different from that in conventional two-dimensional systems.
Due to the chiral nature of charge carriers, it is sensitive not
only to inelastic, phase-breaking scattering, but also to a number
of elastic scattering processes. We study weak localization in
different samples and at different carrier densities, including the
Dirac region, and find the characteristic rates that determine it.
We show how the shape and quality of graphene flakes affect the
values of the elastic and inelastic rates and discuss their physical
origin.

\end{abstract}

\maketitle

The quantum correction to the conductivity of two-dimensional
systems due to electron interference has been studied for more than
twenty years \cite{AltshulerPRB80,Beenaker}. This phenomenon of weak
localization (WL) has become a tool to determine the processes
responsible for electron dephasing due to inelastic electron
scattering or scattering by magnetic impurities
\cite{Beenaker,PierrePRL02}. In this well-established field of
research it comes as a surprise to discover that in a new
two-dimensional system, graphene \cite{NovoselovScience04}, WL does
not follow the standard convention that it is only controlled by
inelastic and spin-flip processes. First attempts to measure WL in
graphene have produced contradictory results that tentatively point
towards this unusual behavior \cite{Morozov,Heersche, Wu}.
Measurements on graphene flakes fabricated by mechanical exfoliation
\cite{Morozov} have shown that in the majority of samples WL is
totally suppressed. In contrast, in a sample fabricated by an
alternative, epitaxial method, WL has been distinctly seen, albeit
at a single (high) carrier density \cite{Wu}.

The theory of WL in graphene \cite{McCannPRL06} predicts a
remarkable feature: it should be sensitive not only to inelastic,
phase breaking processes, but also to different \emph{elastic}
scattering mechanisms \cite{SuzuuraPRL02,MorpurgoPRL06,McCannPRL06}.
The reason for this is that charge carriers in graphene are
\emph{chiral}, that is, they have an additional quantum number
(pseudospin) \cite{Chiral}. Elastic scattering that breaks the
chirality will destroy the interference within each of the two
graphene valleys in $k$-space. Such defects, characterised by the
scattering rate $\tau_s^{-1}$, include surface ripples, dislocations
and atomically sharp defects \cite{Morozov,MorpurgoPRL06}.
Intra-valley WL can also be destroyed by anisotropy of the Fermi
surface in $k$-space, so called `trigonal warping'
\cite{McCannPRL06}, characterised by the rate $\tau_w^{-1}$. There
is one elastic process, however, which acts to restore the
suppressed interference. This is \emph{inter}-valley scattering,
which occurs at a rate $\tau_i^{-1}$ on defects with size of the
order of the lattice spacing $a$. As the two valleys have opposite
chirality and warping, inter-valley scattering is expected to negate
the chirality breaking and warping effects by allowing interference
of carriers from different valleys.

In this work we aim to examine what factors are responsible for the
manifestation of WL in graphene fabricated by mechanical exfoliation
\cite{NovoselovScience04}. We study the magnetoconductivity (MC) in
perpendicular field of several samples with different quality and
dimensions, with the aim to control the relation between the
scattering rates of carriers. These studies are performed at
different carrier densities controlled by a gate voltage $V_g$,
which include densities around the Dirac point at $V_g=0$. This
point is special as about it there is a change of the type of
carrier from electrons to holes and therefore the net carrier
density is zero. The conductivity, however, is seen to remain at a
finite value $\sigma_\mathrm{min}\sim e^2/h$ and not drop to zero
\cite{MinCond}.

\begin{figure}[htb]
\includegraphics[width=.7\textwidth]{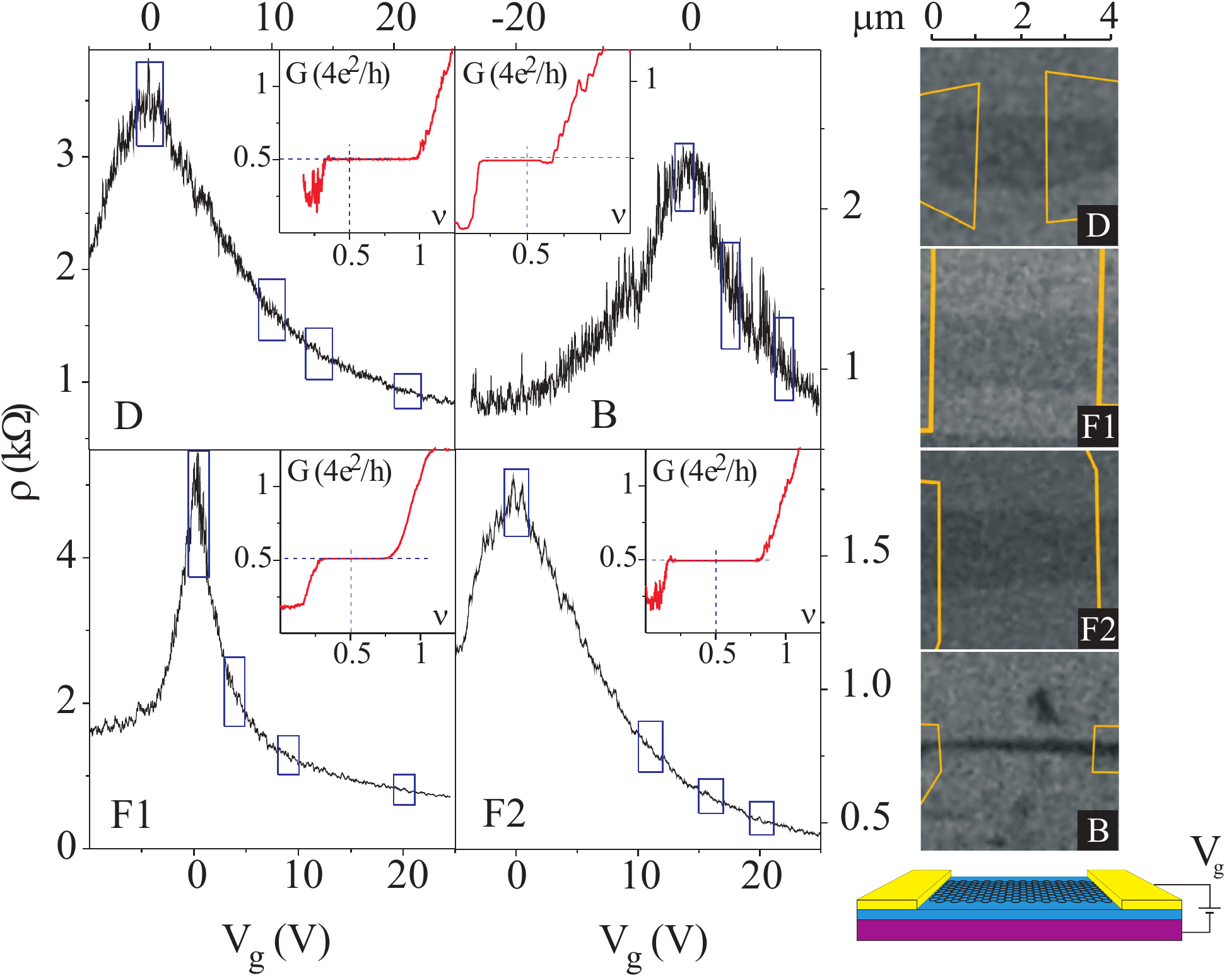}
\caption{Resistivity of graphene flakes as a function of $V_g$ at
$T=$0.25\,K. The mobilities (in $\mathrm{cm^2V^{-1}s^{-1}}$) of the
samples outside the Dirac region: 5100 (D), 7500 (F1), 10000 (F2)
and  8000 (B). The insets show the first quantum Hall plateau, where
filling factor $\nu=nh/4eB$. The right panel shows SEM images of the
samples, where the positions of the contacts are shown as outlines.
The diagram at the foot of this panel shows the graphene sample on
n$^+$Si substrate (purple), covered by 300\,nm SiO$_2$ (blue) and
contacted by Au/Cr (yellow). Control of the carrier density $n$ is
achieved by $V_g$.}\label{fig:one}
\end{figure}

Figure~1 shows the resistivity as a function of $V_g$ of four
samples with different shapes and mobilities: D, F1, F2 and B, each
with a typical peak around the Dirac point. Sample D is a square
flake; F1 and F2 are rectangular with similar width to, but length
larger than D; B is a narrow strip of similar length to F1 and F2
but with much smaller width, Fig.~1 (right). Insets in Fig.~1
demonstrate measurements of the first quantum Hall plateau, which
shows a half-integer step $(0.5\times4e^2/h)$ -- a clear indication
that the samples are single-layer graphene \cite{MinCond}. We wanted
to see what difference the shape of the samples will make to the WL
-- e.g., the narrowest sample B is expected to have the largest
scattering rate $\tau_i^{-1}$, as the edges could produce strong
inter-valley scattering. To understand the relation between the
scattering and the details of the graphene surface, the electrical
measurements have been complemented by atomic force microscope (AFM)
imaging of the sample topography, which have shown the presence of
ripples, and additional rapid folds (ridges) across sample B
(see Supplementary Material below).

\begin{figure}[htb]
\includegraphics[width=.7\textwidth]{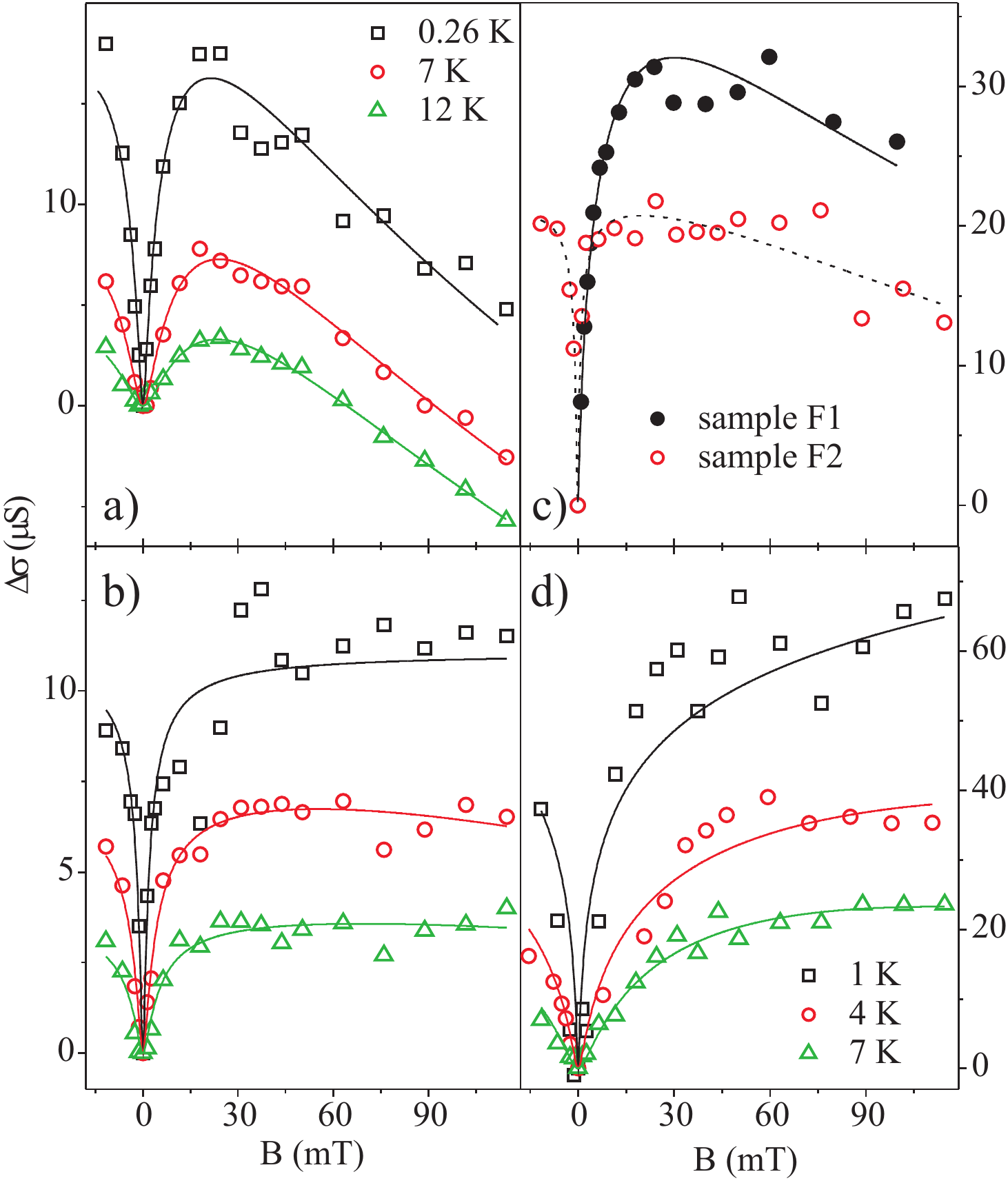}
\caption{(color online) Magnetoconductivity observed in graphene flakes. (a) Dirac
region of sample D, $|V_g|\lesssim 1$\,V, $n \lesssim
7\times10^{10}\,\mathrm{cm^{-2}}$;  (b) sample D, $V_g\simeq14$\,V,
$n\simeq 10^{12}\,\mathrm{cm^{-2}}$ (the legends of (a) and (b) are
the same); (c) samples F1 and F2 at $T=1\,\mathrm{K}$,
$V_g\simeq10$\,V, $n\simeq 7\times10^{11}\,\mathrm{cm^{-2}}$; (d)
sample B, $V_g\simeq11$\,V, $n\simeq
8\times10^{11}\,\mathrm{cm^{-2}}$. Lines are best fits to
Eq.~\ref{eqn:one}.}\label{fig:two}
\end{figure}

In order to study the conductivity correction caused by WL we must
first account for the reproducible conductance fluctuations clearly
seen in Fig.~1. They are also present as a function of magnetic
field $B$ and are caused by the fact that the graphene flakes are
small (comparable to the dephasing length $L_{\phi}$), so that the
average effect of WL is of the same order as the fluctuations which
have the same, quantum interference origin \cite{Beenaker}. We use
the procedure developed in \cite{Gorbachev} in the study of WL in
bilayer graphene: $R(V_g)$ is measured at different $B$ and the
results are averaged over a range $\Delta V_g=2$\,V shown in Fig.~1
by boxes which contain a large number of fluctuations (see Supplementary Material below).
Examples of the obtained average MC, $\triangle\sigma(B)=\langle
\sigma(V_g,B) - \sigma(V_g,0) \rangle_{\Delta V_g}$, are shown in
Fig.~2 for different samples. For the analysis of the results we use
the theory \cite{McCannPRL06} which predicts that the MC is
controlled by several scattering rates, both inelastic
($\tau_\phi^{-1}$) and elastic ($\tau_i^{-1}$, $\tau_{s}^{-1}$,
$\tau_w^{-1}$):
\begin{eqnarray}
\frac{\pi h}{e^2}\cdot\Delta\sigma(B)&=&F\left(\frac{\tau_B^{-1}}
{\tau_{\phi}^{-1}}\right)-F\left(\frac{\tau_B^{-1}}{\tau_{\phi}^{-1}+
2\tau_{i}^{-1}}\right)\nonumber\\
&&-2F\left(\frac{\tau_B^{-1}}{\tau_{\phi}^{-1}+
\tau_{i}^{-1}+\tau_{*}^{-1}}\right)\; .\label{eqn:one}
\end{eqnarray}
Here $F(z)=\ln{z}+\psi{\left(0.5 + z^{-1} \right)}$, $\psi(x)$ is
the digamma function, $\tau_B^{-1}=4eDB/\hbar$, $\tau_\phi^{-1}$ is
the phase-breaking rate and $\tau_*^{-1}=\tau_{s}^{-1}+\tau_w^{-1}$.
(The theory assumes that the momentum relaxation rate $\tau_p^{-1}$
is the highest in the system and comes from the Coulomb charges in
the SiO$_2$ substrate and not atomically sharp defects, so that it
does not affect the carrier chirality.) The first term in Eq. (1) is
responsible for weak localization, while the terms with negative
sign can result in 'anti-localization'.

\begin{figure}[htb]
\includegraphics[width=.7\textwidth]{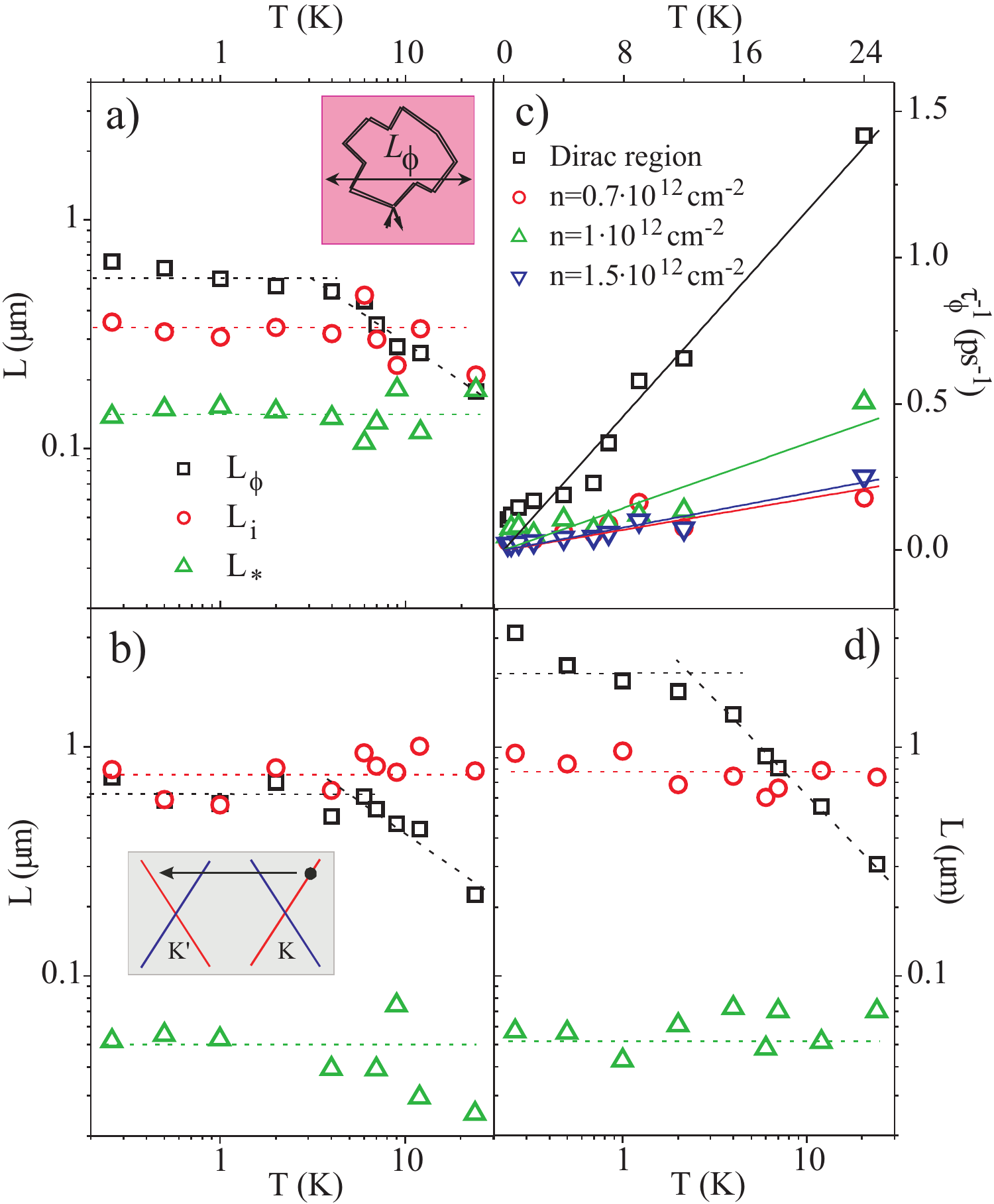}
\caption{(color online) Characteristic lengths responsible for weak
localization; dotted lines are guides to the eye. Sample D: (a) the
Dirac region ($n \lesssim 7\times10^{10}$ cm$^{-2}$) and (b)
electron region ($n\simeq 10^{12}$ cm$^{-2}$); (c) Phase-breaking
rate $\tau_{\phi}^{-1}=D/L_{\phi}^2$ as a function of $T$ for
different $n$. Inset to (a) illustrates the saturation of $L_{\phi}$
at low $T$ due to the sample size. Inset to (b) shows the scattering
process behind the length $L_i$. Sample F2: (d) Temperature
dependence of the characteristic lengths in the electron region
($n\simeq 10^{12}$ cm$^{-2}$).}\label{fig:three}
\end{figure}

We have found that among the different possible relations between
the scattering rates, the following holds in all studied samples:
the intra-valley WL is strongly suppressed due to a large rate
$\tau_*^{-1}$, which approaches $\tau_p^{-1}$; however, WL is
clearly seen in all regions of the carrier density, due to
significant intra-valley scattering, $\tau_i^{-1}
\sim\tau_{\phi}^{-1}$. At the same time, the shape of the MC curves
can be very different as it is controlled by the interplay between
all scattering rates involved, Fig.~2. Comparing two regions of
carrier densities for square sample D (Dirac region (a) and electron
region (b)) one can see that in (a) the curves have a much stronger
downturn, indicating greater importance of the third
(`anti-localising') term in Eq.~\ref{eqn:one} due to smaller rate
$\tau_*^{-1}$. For two geometrically similar samples F1 and F2 in
Fig.~2(c), it is seen that sample F2 (with largest mobility) has a
more rapid increase of the conductivity in smaller field (due to
smaller $\tau_{\phi}^{-1}$) and more rapid downward turn of the
curves at larger fields (due to smaller $\tau_i^{-1}$). For the
narrow sample B, Fig.~2(d), the MC curves do not turn down at all,
indicating a very fast inter-valley rate $\tau_i^{-1}$ and therefore
unimportance of all terms in Eq. 1 apart from the first.

Figure~3 shows the temperature dependence of the characteristic
lengths found from the analysis of the MC by the best fit with
Eq.~\ref{eqn:one}. Figure~3(a,b) compares the results for the Dirac
and electron regions for sample D, where it is seen that $L_{\phi}$
is temperature dependent at high $T$ ($\geq3$ K) but saturates at a
value $L_{\phi}^\mathrm{sat}$ at low temperatures. Figure~3(b,d)
compares the results for samples D and F2, at close values of
carrier densities outside the Dirac region. Sample F2 is about three
times longer, and one can see that $L_{\phi}^\mathrm{sat}$ is larger
in the longer sample $F2$. (In sample B, $L_{\phi}^\mathrm{sat}$ has
also been found to be larger than in sample D.) This clearly implies
that the reason for the saturation is a limitation imposed by the
sample size, and not by scattering from a small, uncontrolled number
of magnetic impurities \cite{PierrePRL02}.

In the Dirac region $L_{\phi}^\mathrm{sat}$ has been found to have a
smaller value than at larger carrier densities. (It is interesting
to note that the narrow sample, B, shows the biggest decrease of
$L_{\phi}^\mathrm{sat}$ in the Dirac region, while the square
sample, D, the smallest.) This decrease can be related to the
inhomogeneity of the sample at low carrier densities.  It can result
in formation of electron--hole puddles, so that at $V_g=0$ there are
equal densities of electrons and holes and not zero density of each
type of carrier. Inhomogeneity can modify the geometry of conducting
paths and decrease the effective dimensions of the sample, resulting
in a smaller value of $L_{\phi}^\mathrm{sat}$.

The temperature dependence of $L_{\phi}$ contains information about
the inelastic mechanism responsible for the dephasing of charge
carriers. There are suggestions that electron-electron (\emph{e-e})
interaction, the main mechanism of dephasing at low $T$, is
different in graphene compared with other systems \cite{e-eInt}. To
examine this we have analyzed the $T-$dependence of the dephasing
rates found from analysis of the WL. Figure~3(c) shows the
phase-breaking rate in different density regions of sample D. (To
find $\tau_{\phi}^{-1}$ we use the relation $L_{\phi}=(D
\tau_{\phi})^{1/2}$, where the diffusion coefficient $D=v_Fl/2$ is
determined from the mean free path $l=h/2e^2k_F \rho$. For the Dirac
region, where the puddles can be formed, the value of the Fermi
wavenumber $k_F$ (inside the puddle) is simply estimated at the
boundary of the region where the MC is studied, $|V_g|=1$ V,
Fig.~1.) Our results show that electron dephasing rate obeys the
usual, linear $T-$dependence for \emph{e-e} scattering in the `dirty
limit', $T\tau_p<1$ \cite{AltshulerPRB80}: $\tau_{\varphi}^{-1} =
\beta k_BT\ln{g} /\hbar g$, where $g=\sigma h/e^2$. (In our samples
the parameter $T\tau_p$ varies from 0.002 to 0.4 in the studied
temperature range 0.25--25 K.) The empirical coefficient $\beta$ is
found to be between 1 and 2 in all studied regions, Fig.~3(c), and
all samples. Therefore, we can conclude that while electron
interference in graphene is significantly different from other
systems, \emph{e-e} interaction does not show unconventional
behavior.

In addition to the analysis of the WL, we have also analyzed
conductance fluctuations as a function of $B$ and $V_g$ using
standard relations in terms of $L_{\phi}$ \cite{Beenaker}. This
analysis has given values of $L_{\phi}$ close to those found from
the analysis of WL (see Supplementary Material below).

Now let us discuss the behavior of the elastic, inter-valley length
$L_i$ which we have verified to be essentially $T-$independent in
all samples. In samples D, F1 and F2 the found $L_i$ is comparable
to the width of the samples (approximately half the width). This
means that, indeed, the sample edges make significant contributions
to inter-valley scattering. This is consistent with the fact that
the narrowest sample B has shown the smallest value of $L_{i}$.
However, the value of $L_i$ for sample B is about three times
smaller than the sample width. This can be due to the presence of
rapid ridges of height $\sim1.5\,\mathrm{nm}$ observed in this
sample by AFM, Fig.~4(a). They are separated by a distance smaller
than the sample width and can be another source of inter-valley
scattering. This suggests that the inter-valley scattering is
controlled not only by the edges but also by the defects in the
inner part of the sample.

We have found that the intra-valley scattering length $L_*$ is much
smaller than the inter-valley length $L_i$, Fig.~3, and approaches
the mean free path. There are several possible mechanisms that can
be responsible for the observed large intra-valley scattering rate
and resulting strong suppression of WL in one valley (see Supplementary Material below).
Scattering by \emph{atomically-sharp defects} cannot explain this:
such scattering is also a source of strong inter-valley scattering,
so that $L_i$ and $L_*$ would be comparable if this mechanism was
dominant. The smaller value of $L_*$ in experiment must therefore be
due to an additional scattering rate which affects $L_*$ but not
$L_i$: from \emph{warping} \cite{McCannPRL06}, or from the defects
of the crystal structure that are large on the atomic scale
\cite{Morozov,MorpurgoPRL06}. Estimation of the expected
$\tau_w^{-1}$ using theory  \cite{McCannPRL06}  gives a value of
$\tau_w^{-1}\leq 0.3$ ps$^{-1}$ which is much smaller than the
experimental $\tau_*^{-1}\sim 10$ ps$^{-1}$. Therefore, the reason
for small $L_*$ could lie in the defects of the crystal structure of
graphene flakes: \emph{ripples} and \emph{dislocations}. (The strain
in the lattice induced by such defects acts as a source of effective
magnetic field that can destroy WL.)  As the dislocation core is
also a source of inter-valley scattering, their separation can be
estimated from the known value of $L_i$. This value is much larger
than the dislocation separation $\xi\sim$50\,nm required to explain
the small value of $L_*$. (This value of $\xi$ is obtained using the
relation $\tau_{s}^{-1}=v_F/k_F\xi^2$ \cite{MorpurgoPRL06}.) The
effect of ripples on the graphene surface is also negligible in our
samples, if we use the estimation of this effect from
\cite{Morozov}. The roughness of our samples found from AFM
measurements, Fig.~4(b,c), is $h\simeq 0.3$\,nm and $d\simeq 10$\,nm
(in agreement with \cite{Ishigami}), which gives for a typical
$L_\phi\approx 1\,\mathrm{\mu m}$ an effective magnetic field
$B_\mathrm{eff}\sim 1$ mT. This is a small correction to the real
fields used in experiment, Fig.~2. There is one more mechanism that
can introduce an asymmetry in the crystal structure and hence break
the chirality of carriers: a potential gradient coming from charged
impurities in the substrate. Our estimation of its effect using the
approach of \cite{MorpurgoPRL06} has also given a negligible result
(see Supplementary Material below). Therefore, the obtained values of the scattering rate
$\tau_*^{-1}$ are much higher than those predicted by existing
models, and more detailed theories of the mechanisms of intra-valley
suppression of WL in graphene are required.

\begin{figure}[htb]{}
\includegraphics[width=0.7\textwidth]{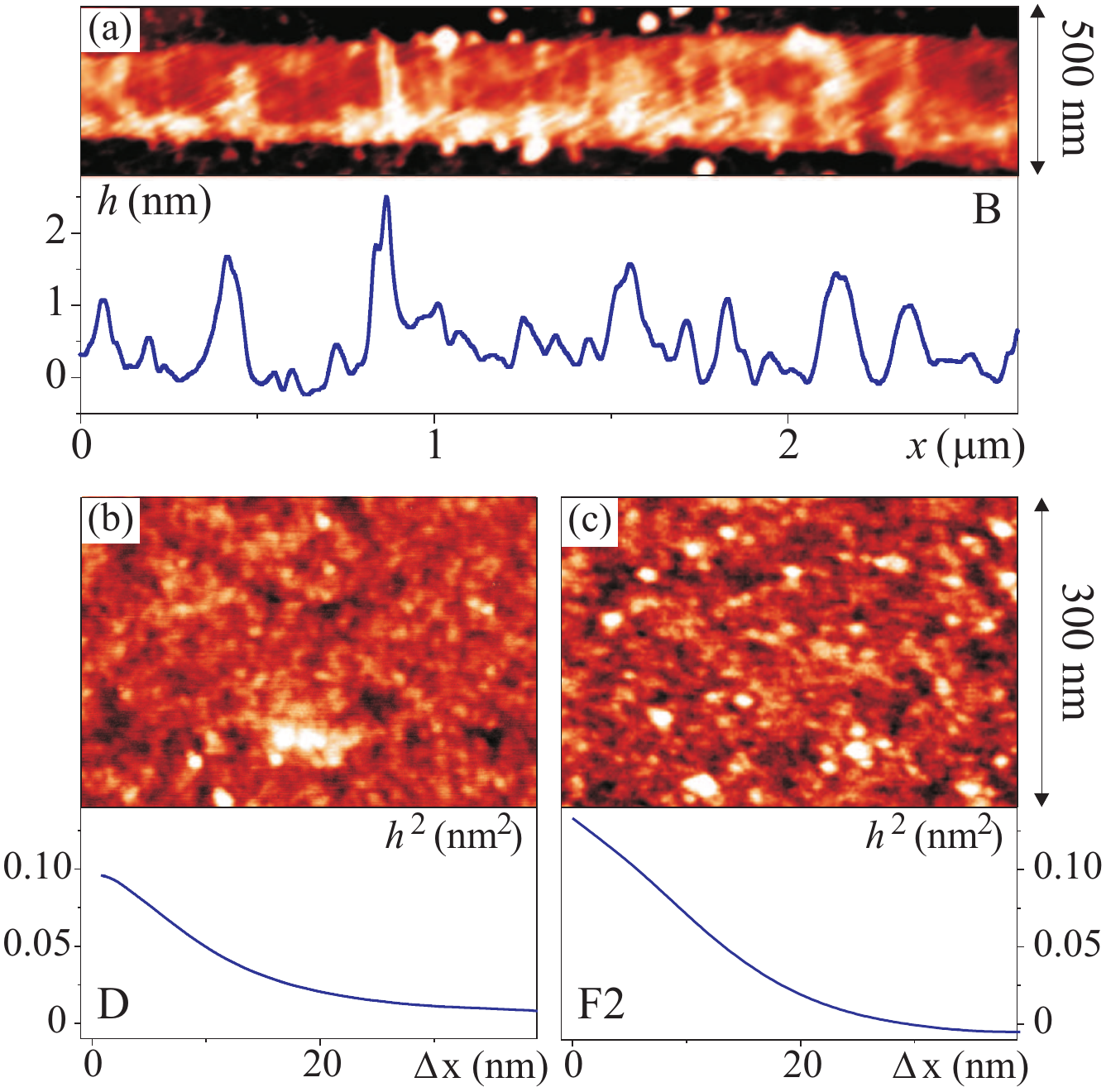}
\caption{Atomic force microscope images of graphene surfaces. Below
the image of sample B (a) is the surface profile averaged over the
width of the sample. Below the images of the topography of samples D
(b) and F2 (c) are the corresponding autocorrelation functions of
the surface roughness.}\label{fig:four}
\end{figure}

In summary, we have shown that the weak  localization correction in
graphene exists at all studied carrier densities, including the
Dirac region. Its manifestation is determined by the interplay of
inelastic and elastic scattering mechanisms, which makes WL a
sensitive tool to detect the defects responsible for inter-valley
scattering and chirality breaking. We show that, in spite of a
strong intra-valley suppression of WL, the quantum interference
correction to the conductivity is clearly seen due to significant
inter-valley scattering. Total suppression of WL is only possible in
experiments where inter-valley scattering is negligible, i.e. in
very large samples without sharp defects in the bulk.

We gratefully acknowledge stimulating discussions with E.~McCann,
V.~V.~Cheianov, F.~Guinea and V.~I.~Fal'ko, and thank B.~Wilkinson
for assistance at an early stage of the experiments.

\clearpage \setcounter{figure}{0}

\begin{center}
\textbf{\Large Weak localisation in graphene flakes: Supplementary
material}

\bigskip

F.~V.~Tikhonenko, D.~W.~Horsell, R.~V.~Gorbachev, and
A.~K.~Savchenko

\smallskip

\textit{School of Physics, University of Exeter, Stocker Road,
Exeter, EX4 4QL, U.K.}
\end{center}

\bigskip

\bigskip

\noindent\textbf{Samples}

Samples were manufactured using the method of mechanical exfoliation
of highly-oriented pyrolytic graphite devised in
\cite{NovoselovScience04}, on a n$^{+}$Si/SiO$_2$ substrate with
oxide layer of thickness $t=300$\,nm. Lithographically defined Au/Cr
contacts were subsequently made to each flake. Resistance
measurements were carried out in the temperature range from $0.25$
to $25\,\mathrm{K}$ using a standard lock-in technique with
$1\,\mathrm{nA}$ driving current. Samples B, D, F1 are two-terminal
and F2 is four-terminal (the additional contacts were used to
account for the contact resistance). The concentration of carriers
(electrons $n$ and holes $p$) in graphene is determined by the
capacitance between the graphene and n$^{+}$Si substrate:
$e(p-n)=(\epsilon \epsilon_0 /t)V_g$. There was a small
unintentional doping of the samples leading to a shift in gate
voltage ($\sim5\,\mathrm{V}$) of the position of the resistance peak
with respect to $V_g=0$, which has been accounted for in the main
text. The graphene--Au/Cr contact resistance has been found from the
deviation of the height of the quantum Hall plateau from the
expected value of $2e^2/h$ (see insets to Fig. 1 of main text). The
values of the contact resistance for samples F1 and D are about
$\sim$100\,$\Omega$ and $\sim$600\,$\Omega$ for sample B.

\bigskip

\noindent\textbf{Averaging procedure and analysis of
magnetoconductance}

\begin{figure}[htb]{}
\includegraphics[width=0.7\columnwidth]{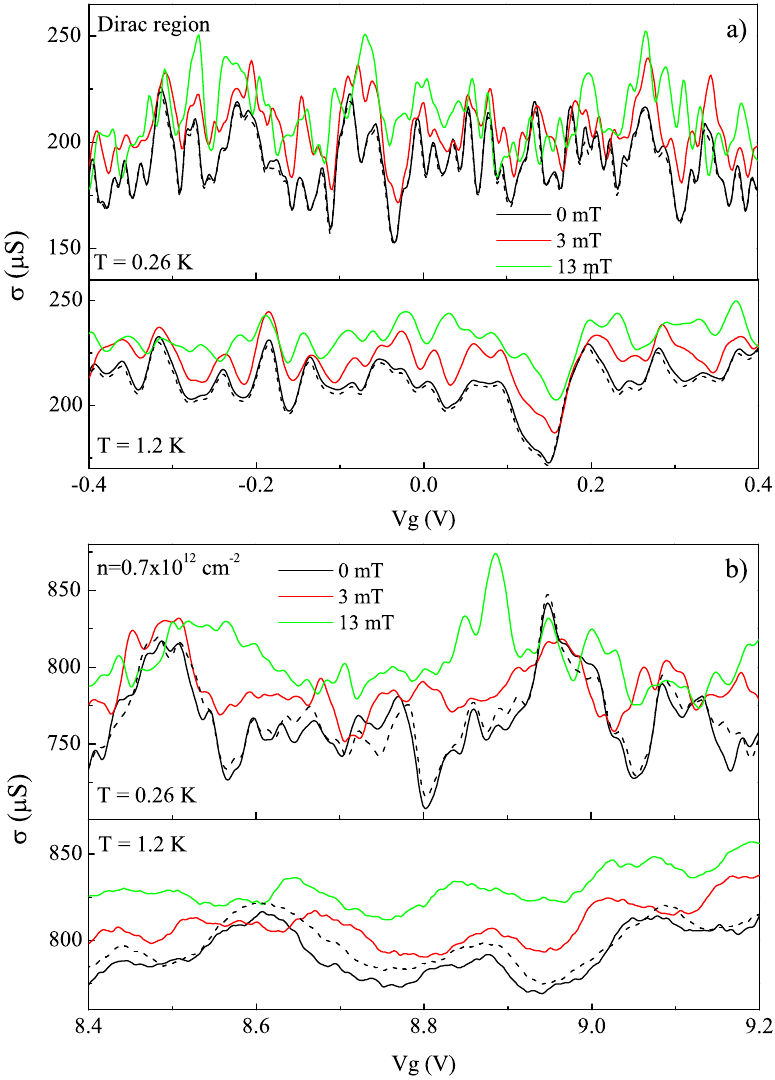}
\caption{Illustration of the averaging procedure of the
magnetoconductance of sample F1 in two density regions at  two
temperatures (only a fraction of $\Delta V_g$ is shown here): (a)
Dirac region, (b) electron region. Dotted lines show repeated sweeps
at $B=0$.}\label{fig:WLmethod}
\end{figure}

A method of effective averaging is important in small-sized samples
to remove the influence of mesoscopic fluctuations, as without it
one can get contradictory results for the magnetoconductance (MC).
(If we attempt to measure $\Delta\sigma(B)$ at different $V_g$, the
character of the MC depends on the specific point in $V_g$ at which
it is measured). Figure~\ref{fig:WLmethod} shows how the averaging
is performed. For each temperature the conductivity of the sample as
a function of the gate voltage is first measured across a
$2\,\mathrm{V}$ range at incremental values of the magnetic field.
Then the curve at zero magnetic field is subtracted from each curve
and the resulting difference is averaged across the $2\,\mathrm{V}$
gate voltage range. One can see from Fig. 1 the average increase of
$\langle\Delta\sigma\rangle_{\Delta V_g}$ with magnetic field. These
averaged values of the MC are plotted as a function of $B$ in Figure
2 of the main text.

The perturbation theory of weak localization (WL) is applicable at
$k_Fl\gg1$ (a diffusive metal). In our samples $k_Fl$, found from
the conductivity $\sigma=2e^2(k_Fl)/h$, varies in the range 3--30,
with the smallest values in the Dirac region: 4, 3, 8 and 6 for
samples D, F1, F2 and B, respectively. Another limitation for the
application of the diffusive theory of WL is $B\lesssim B_{\mathrm
{tr}}$, where the `transport' magnetic field is found from the
condition $L_B=(\hbar/eB)^{1/2}\approx l$. This limits the range of
magnetic fields where we perform the analysis to
$B\leqslant100$\,mT.

For the narrowest sample B, the dephasing length is larger than its
width and therefore the 1D theory of WL \cite{McCannPRL06} should be
used in the analysis of its MC in small fields. However, at fields
where $L_B<W$ ($W$ is the width of the sample) i.e. at $B>7$ mT, the
2D theory becomes applicable. As the bulk of the data is obtained in
this range of the field, we have used 2D theory (Eq.~1 in main text)
to analyse the MC.

\bigskip

\noindent\textbf{Comparison of characteristic lengths and times}

\begin{figure}[htb]{}
\includegraphics[width=0.7\columnwidth]{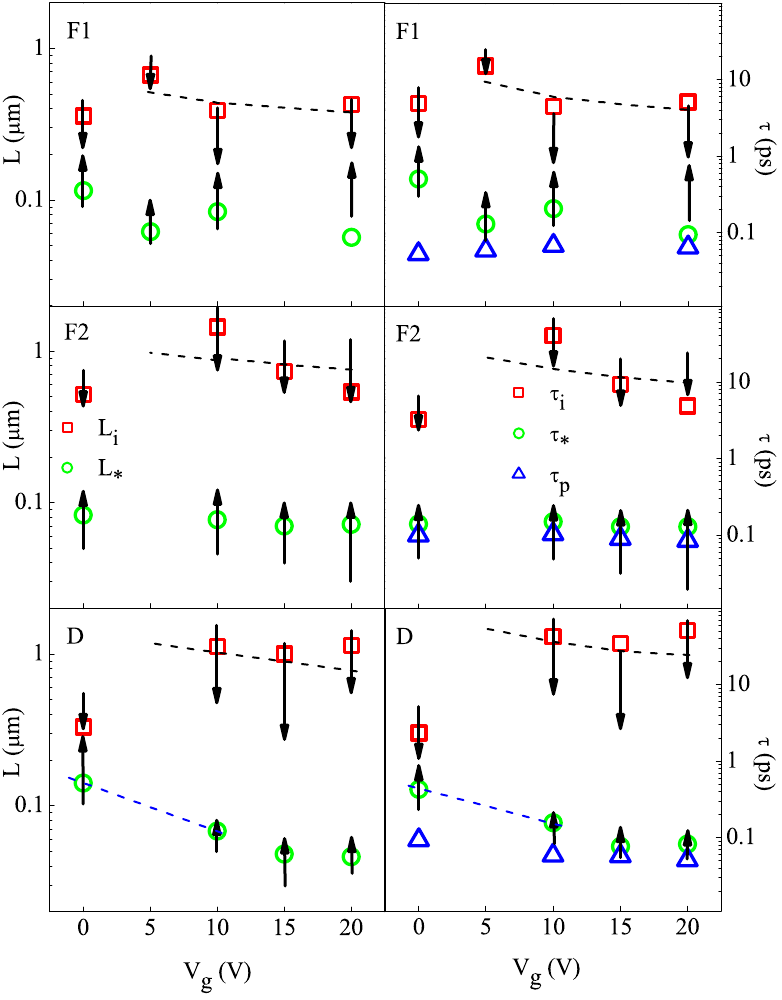}
\caption{Comparison of characteristic lengths and times for samples
F1, F2 and D at different carrier densities.}\label{fig:ltau}
\end{figure}

Figure~\ref{fig:ltau} shows for samples F1, F2 and D a comparison of
the length $L_*$ with length $L_i$, as well as the values of the
corresponding times $\tau_*$ and $\tau_i$ (using
$L_x=(D\tau_x)^{1/2}$) for different carrier densities. We emphasise
that in the analysis of the MC the value of $L_*$ is closely linked
to that of $L_i$. In Eq.~1 the second and third terms have the same
sign, therefore by a slight increase of one of them and a
corresponding decrease of the other, one can get a similar agreement
with experiment. Figure~\ref{fig:ltau} shows not only the values
found from the best fit (the higher $B$-region being most sensitive
to these two parameters) but also the synchronous variation allowed
in these values while retaining a good fit, indicated by arrows. In
spite of the variations, there are several trends seen in the
figure. First, the value of $L_i$ is always significantly larger
than $L_*$ and somewhat larger in the better quality sample F2.
Second, there is a decrease of $L_i$ with increasing carrier
density, although its value is smaller in the Dirac region. Finally,
there is a decrease of $L_*$ when the carrier density is increased
above the Dirac region. The dashed curves in Fig.~2 indicate the
expected decrease of $L_i$ and $\tau_i$ if the scattering rate is
proportional to the density of states, which increases linearly with
the Fermi energy $\epsilon_F\propto V_g^{1/2}$.

\bigskip

\noindent\textbf{Estimations of the effects suppressing WL in a
single valley}

\noindent\emph{Trigonal warping}

According to \cite{McCannPRL06} the breaking of the time-reversal
symmetry in one valley can occur due to the suppression of
backscattering by the trigonal warping of the Fermi surface. The
trigonal warping rate is
\[
\tau_w^{-1} = 2\tau_p(\mu \epsilon_F^2/\hbar v_F^2)^2\; ,
\]
where $\tau_p$ is the momentum relaxation time, $v_F\approx
10^6\,\mathrm{ms^{-1}}$ is the Fermi velocity and $\mu$ is the
structural parameter equal to $\mu = {\gamma_0 a^2}/{8\hbar^2}$.
Here $\gamma_0\approx 3$\,eV is the nearest-neighbour hopping energy
and $a\approx 0.26$\,nm is the lattice constant in graphene. For the
typical parameters in our samples we obtain $\tau_w^{-1}\approx
0.001$\,ps$^{-1}$ for the Dirac region ($\epsilon_F\approx 30$\,meV,
$\tau_p\approx 0.1$\,ps) and $\tau_w^{-1}\approx 0.3$\,ps$^{-1}$ for
the highest measured concentration ($\epsilon_F\approx 130$\,meV,
$\tau_p\approx 0.05$\,ps). Trigonal warping of the Fermi surface is
therefore a very weak effect compared to other intra-valley
scattering mechanisms and cannot be the main reason of the strong
chirality-breaking observed in our experiments
($\tau_*^{-1}\approx\tau_p^{-1}$).

\bigskip

\noindent\emph{Dislocations}

Another possible mechanism of chirality breaking in the graphene
sheet is dislocations in the honeycomb lattice \cite{MorpurgoPRL06}.
If the trajectory of a quasiparticle goes near the core of a
dislocation it leads to a change of the phase due to the induced
strain. For randomly distributed dislocations the scattering rate
related to this mechanism is
\[
\tau_\mathrm{gauge}^{-1}\approx\frac{v_F}{k_F\xi^2}\; ,
\]
where $\xi$ is the average distance between dislocations
\cite{MorpurgoPRL06}. In order to obtain the experimentally found
chirality-breaking rate $\tau_{*}^{-1}\approx$ 10 -- 20\,ps$^{-1}$
the distance $\xi$ should be about $15-50$\,nm. However, the cores
of the dislocations should also cause inter-valley scattering, which
is why this estimation is in contradiction with the relatively large
value of the inter-valley scattering length ($L_i\approx 1 \mu$m)
observed experimentally.

\bigskip

\noindent\emph{Ripples}

As proposed in \cite{Morozov}, ripples in the graphene layer on a
silica substrate can lead to suppression of weak localization
because of the effective magnetic field generated by strain of the
interatomic bonds. The vector potential corresponding to a single
ripple with diameter $d$ and height $h$ is \cite{Morozov}:
\[
A = \frac{\gamma_0 \left|\nabla{h}\right|^2}{ev_F}\; ,
\]
where $\nabla{h}\approx h/d$. The flux through one ripple is
$\Phi=\oint \textbf{A}\cdot\mathrm{d}\textbf{l}\approx Ad$ and
$\Phi=\int \textbf{B}\cdot \mathrm{d}\textbf{S}\approx Bd^2$,
therefore the magnetic field associated with one ripple is
\[
B\approx \frac{A}{d}=\frac{\gamma_0}{ev_F}\frac{h^2}{d^3}\; .
\]
Since the curvature vector of a ripple is random, the resulting
magnetic field through the area limited by the dephasing length
$L_\phi$ and containing $N \approx L_{\phi}^2/d^2$ ripples should be
averaged as follows:
\[
B_\mathrm{eff}=\frac{B}{\sqrt{N}} =
\frac{\gamma_0}{ev_FL_{\phi}}\left(\frac{h}{d}\right)^2\;.
\]
The roughness of the graphene sheet found from AFM measurements is
about $0.3$\,nm and the size of the features is about $10$\,nm. This
gives a value for the magnetic field associated with one ripple
$B\sim 0.1$\,T. For our typical value of $L_{\phi}\sim 1\,\mu$m the
effective magnetic field is then $B_\mathrm{eff}\sim 1$\,mT. Since
suppression of the quantum interference requires a magnetic field
$B_\mathrm{eff}>B_\mathrm{tr}\sim 0.1$\,T, the estimated value is
too small to destroy the localization effect. The random magnetic
field can only introduce an uncertainty in the value of $B$, Fig.~2
of the main text, comparable to the accuracy to which the field is
set by the power supply.

\bigskip

\noindent\emph{Potential gradients}

The last mechanism which can produce the breaking of time-reversal
symmetry is a gradient of potential coming from the charged
impurities in the substrate. A potential gradient leads to a
distortion of the dispersion curve of a single valley and hence
breaks the valley symmetry. As shown in \cite{MorpurgoPRL06} the
resulting scattering rate can be estimated as
\[
\tau_\mathrm{grad}^{-1}\approx\tau_p^{-1}\left(k_Fa\right)^2\; .
\]
In order to get $\tau_\mathrm{grad}^{-1}\approx\tau_p^{-1}$ one
should have $k_Fa\approx 1$. This corresponds to the carrier density
$n=k_F^2/\pi\approx 5\cdot10^{14}$\,cm$^{-2}$, which is two orders
of magnitude higher than the densities studied in the experiment.

We conclude from these calculations that all existing estimations
for the chirality-breaking scattering rates are not sufficient to
explain our experimental results.

\bigskip

\noindent\textbf{Scanning probe microscopy studies}

The atomic force microscope used in this work was an Ntegra Aura
from NT-MDT. We used non-contact tips NSG01 with resonance of
$150\,\mathrm{Hz}$ at an amplitude of $\lesssim40\,\mathrm{nm}$. To
obtain high resolution in the $xy$-plane `diamond-like carbon'
coated tips with curvature radius $1-3\,\mathrm{nm}$ were used; tip
convolution therefore limited feature resolution to this scale. To
remove the influence of the water layer present on the silica
substrate all measurements were performed in an atmosphere of dry
nitrogen at $3\,\mathrm{mbar}$, giving a tip resonance quality
factor $Q\approx1000$. The noise in the $z$-scale (height) is of the
order $0.02\,\mathrm{nm}$ measured on pure graphite and silica with
the AFM operating with acoustic and vibrational isolation.

\begin{figure}[htb]{}
\includegraphics[width=0.7\columnwidth]{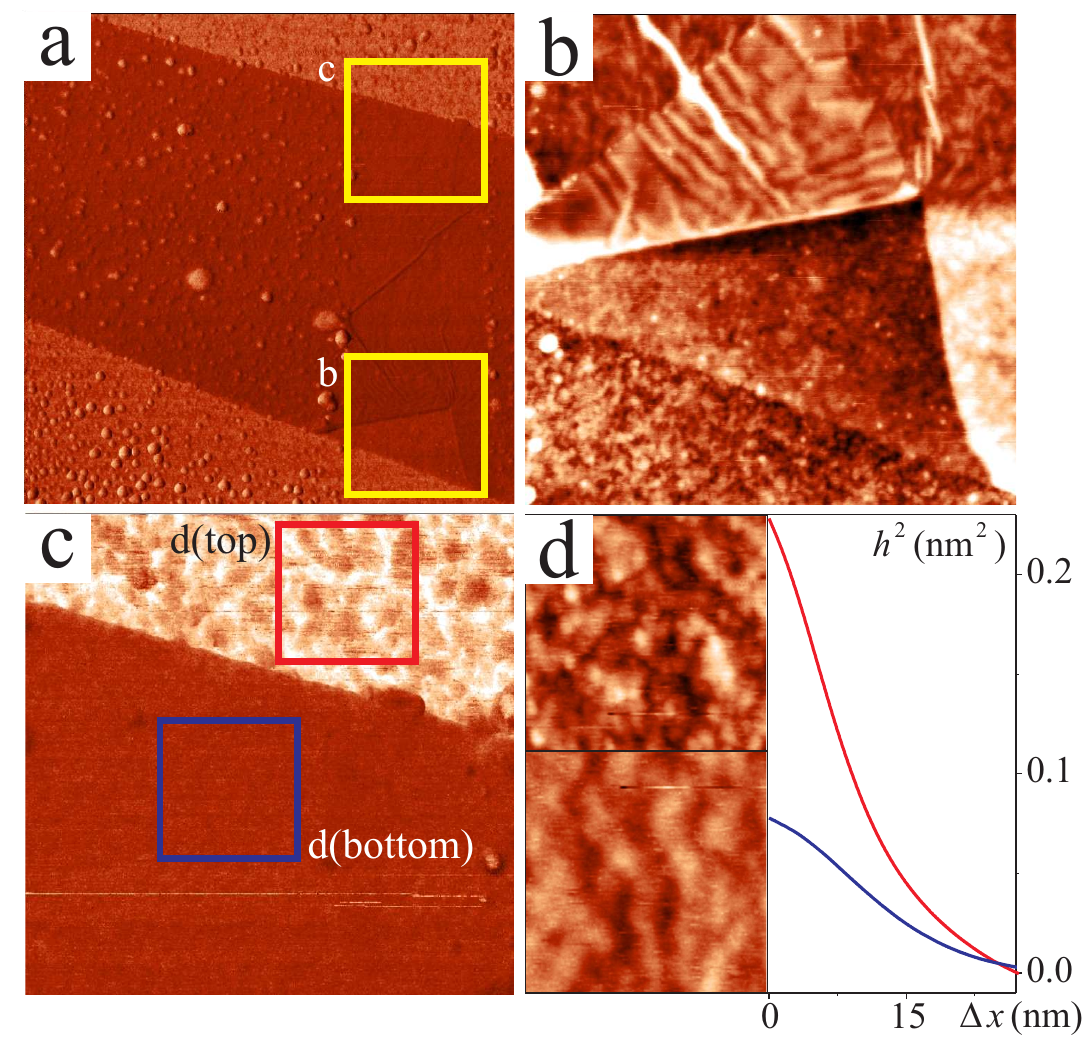}
\caption{AFM measurements of sample F2. (a) Phase contrast image
where the PMMA droplets on the left and cleaned area on the right
can be seen, with two regions in the clean area highlighted by
boxes. (b) Magnified topographic image of box `(b)' showing the torn
edge of the graphene flake. (c) Magnified phase contrast image of
box `(c)' with silica (top) and graphene (bottom). (d)
Autocorrelation analysis of the roughness in the boxes highlighted
in (c), with insets of silica (top) and graphene topography
(bottom). Scan (a) is $3\,\mathrm{\mu m}$ size and the phase change
at the graphene--silica boundary is $2^\circ$. Scans (b)--(c) have
the same $0.8\,\mathrm{\mu m}$ size. In (b) the colour-scale varies
over $4\,\mathrm{nm}$.}\label{fig:threeS}
\end{figure}

We found that the surfaces of the silica and the graphene after
lithographic processing were covered in droplets of PMMA with height
$\approx2\,\mathrm{nm}$, similar to the findings of
\cite{Ishigami}. They reduced the image quality and also made
determination of the step edge between graphene and silica
difficult. To obtain the scans of clean graphene shown in Fig.~4 of
the main text we mechanically cleaned the surfaces.
Figure~\ref{fig:threeS}(a) shows a phase contrast image of sample F2
where both the PMMA droplets and a cleaned area are seen.

To understand the extent to which the PMMA droplets exist under the
flake (due to the lithographic process of depositing location
markers prior to the deposition of the graphene flake), we
introduced a tear and fold into the sample F2 as seen in
Fig.~\ref{fig:threeS}(b). We see first that the surface under the
flake is indeed free from PMMA droplets and therefore the topography
of the flake is only influenced by the silica roughness. (This
conclusion was also confirmed by similar measurements on other
flakes). Having a flake fold allows us to determine better the
thickness of the flake, by measuring the step height between two
graphene areas (as opposed to measurements of the step height
between silica and graphene which always give a larger value of the
step, $\sim1\,\mathrm{nm}$).  We find that the thickness of the
flake is $<0.5\,\mathrm{nm}$, which confirms that the flake is a
monolayer (supporting the results of the quantum Hall measurements
discussed in the main text). An interesting result from the tear is
that the graphene flake has a tendency to form larger ripples when
detached from the silica surface, with a ripple height
$\sim0.5\,\mathrm{nm}$ and width $20\,\mathrm{nm}$. (The roughness
of the flake on the substrate is $\sim 0.3$\,nm, see the main text.)
When comparing the surface roughness of silica and graphene,
Fig.~\ref{fig:threeS}(c,d), we see that the surface height variation
on the clean silica surface is $\sim60\%$ larger than on the
graphene, i.e.~graphene significantly smoothes out the substrate
roughness.


\begin{thebibliography}{99}
\bibitem{AltshulerPRB80} B.~L.~Altshuler, D.~Khmel'nitzkii, A.~I.~Larkin, and P.~A.~Lee, Phys. Rev. B \textbf{22}, 5142
(1980); G.~Bergman, Phys. Rep. \textbf{107}, 1 (1984).
\bibitem{Beenaker} C.~W.~J.~Beenakker, and  H.~Van Houten, Solid State Physics \textbf{44}, 1  (edt. by H. Ehrenreich and D. Turnbull, Academic Press Inc., San Diego, 1991).
\bibitem{PierrePRL02} F. Pierre, A. B. Gougam, A. Anthore, H. Pothier, D. Esteve, and N. O. Birge, Phys. Rev. B \textbf{68} 085413 (2003).
\bibitem{NovoselovScience04} K.~S.~Novoselov, \textit{et al.}, Science \textbf{306}, 666 (2004).
\bibitem{Morozov} S.~V.~Morozov, \textit{et al.}, Phys. Rev. Lett. \textbf{97}, 016801 (2006).
\bibitem{Heersche} H.~B.~Heersche \textit{et al.}, Nature \textbf{446}, 56 (2007).
\bibitem{Wu} X.~Wu, X.~Li, Z.~Song, C.~Berger, and W.~A.~de Heer, Phys. Rev. Lett. \textbf{98}, 136801 (2007).
\bibitem{McCannPRL06} E.~McCann, \textit{et al.}, Phys. Rev. Lett. \textbf{97}, 146805 (2006).
\bibitem{SuzuuraPRL02} H.~Suzuura, and T.~Ando, Phys. Rev. Lett. \textbf{89}, 266603 (2002).
\bibitem{MorpurgoPRL06} A.~F.~Morpurgo, and F.~Guinea, Phys. Rev. Lett. \textbf{97}, 196804 (2006).
\bibitem{Chiral} T.~Ando, T.~Nakanishi, and R.~Saito, J. Phys. Soc. Japan \textbf{67}, 2857 (1998).
\bibitem{MinCond} A.~K.~Geim, and K.~S.~Novoselov, Nature Mat. \textbf{6}, 183 (2007).
\bibitem{Gorbachev} R.~V.~Gorbachev, F.~V.~Tikhonenko, A.~S.~Mayorov, D.~W.~Horsell and A.~K.~Savchenko, Phys. Rev. Lett. \textbf{98}, 176805 (2007).
\bibitem{e-eInt} J.~Gonzalez, F.~Guinea, and M.~A.~H.~Vozmediano, Phys. Rev. Lett. \textbf{77}, 3589 (1996); J.~Gonzalez, F.~Guinea, and M.~A.~H.~Vozmediano,
 Phys. Rev. B \textbf{59}, R2474 (1999); S.~Das Sarma, E.~H.~Hwang and W-K.~Tse, Phys. Rev. B \textbf{75}, 121406(R) (2007); M.~Polini, R.~Asgari, Y.~Barlas, T.~Pereg-Barnea and A.~H.~MacDonald, arXiv:0704.3786.
\bibitem{Ishigami} M.~Ishigami, J.~H.~Chen, W.~G.~Cullen, M.~S.~Fuhrer, and E.~D.~Williams, Nano Letters \textbf{7}, 1643 (2007).
\end{thebibliography}
\end{document}